\documentclass[twocolumn,showpacs,preprintnumbers,amsmath,amssymb,prl,floatfix]{revtex4}
\usepackage{rotating}
\usepackage{graphicx}
\usepackage{dcolumn}
\usepackage{bm}
\usepackage{multirow}
\usepackage{natbib}

\begin{document}

\title{Limits on an Energy Dependence of the Speed of Light from \\a
  Flare  of the
  Active Galaxy PKS 2155-304}

\author{F. Aharonian$^{1,13}$ }
\author{ A.G.~Akhperjanian$^{2}$ }
\author{ U.~Barres de Almeida$^{8}$} \thanks{supported by CAPES Foundation, Ministry of
Education of Brazil} 
\author{ A.R.~Bazer-Bachi$^{3}$ }
\author{Y.~Becherini$^{12}$}
\author{ B.~Behera$^{14}$ }
\author{ M.~Beilicke$^{4}$ }
\author{ W.~Benbow$^{1}$ }
\author{ K.~Bernl\"ohr$^{1,5}$ }
\author{ C.~Boisson$^{6}$ }
% ...

\footnotesize
\affiliation{$^{1}$
Max-Planck-Institut f\"ur Kernphysik, P.O. Box 103980, D 69029
Heidelberg, Germany
}\affiliation{$^{2}$
 Yerevan Physics Institute, 2 Alikhanian Brothers St., 375036 Yerevan,
Armenia
}\affiliation{$^{3}$
Centre d'Etude Spatiale des Rayonnements, CNRS/UPS, 9 av. du Colonel Roche, BP
4346, F-31029 Toulouse Cedex 4, France
}\affiliation{$^{4}$
Universit\"at Hamburg, Institut f\"ur Experimentalphysik, Luruper Chaussee
149, D 22761 Hamburg, Germany
}\affiliation{$^{5}$
Institut f\"ur Physik, Humboldt-Universit\"at zu Berlin, Newtonstr. 15,
D 12489 Berlin, Germany
}\affiliation{$^{6}$
LUTH, Observatoire de Paris, CNRS, Universit\'e Paris Diderot, 5 Place Jules Janssen, 92190 Meudon,
France
Obserwatorium Astronomiczne, Uniwersytet Ja
}
% ...
\affiliation{$^{8}$
University of Durham, Department of Physics, South Road, Durham DH1 3LE,
U.K.
}
% ...
\affiliation{$^{12}$
Astroparticule et Cosmologie (APC), CNRS, Universite Paris 7 Denis Diderot,
10, rue Alice Domon et Leonie Duquet, F-75205 Paris Cedex 13, France
\thanks{UMR 7164 (CNRS, Universit\'e Paris VII, CEA, Observatoire de Paris)}
}\affiliation{$^{13}$
Dublin Institute for Advanced Studies, 5 Merrion Square, Dublin 2,
Ireland
}\affiliation{$^{14}$
Landessternwarte, Universit\"at Heidelberg, K\"onigstuhl, D 69117 Heidelberg, Germany
}
% ...

\begin{abstract}
\normalsize
In the past few decades, several models have predicted an energy-dependence
of the speed of light in the context of quantum gravity. For
cosmological sources such as active galaxies,
this minuscule effect can add up to
measurable photon-energy dependent time lags. In this paper a search for such time
lags during the H.E.S.S. observations of the exceptional very high energy flare of the active galaxy
PKS 2155-304 on 28 July in 2006 is presented. Since no significant time lag is found,  lower
limits on the energy scale of speed of light modifications are
derived.
\end{abstract}
\maketitle
\normalsize

Albert Einstein's postulate \textit{``that light is always propagated in empty space with a definite velocity c which
is independent of the state of motion of the emitting body''}
\cite{EINSTEIN} is one of the pillars of modern physics.
Modification of this postulate would have far-reaching consequences
for our understanding of nature, it is therefore important to
constantly improve the verification of its validity.
Particularly in the past few decades, a possible energy dependence of the
speed of light has been predicted in the framework of quantum gravity
models \cite{LOOP1,STRING5,DSR}
and effective field theory \cite{EFT}, leading to deviations from this postulate (for reviews
see \cite{AMELINONATURE,OVERVIEW1,FARAKOS}). The speed of light
modifications have different functional dependencies on the
photon energy and helicity in different models. Predictions usually entail free parameters
such as the relevant mass scale. However, it is commonly expected that this
modification should appear at energies of the order of the Planck
energy ($E_{\rm P} = 1.22 \times 10^{19}$~GeV). 
For energies much smaller
than the Planck energy, a series
expansion is therefore expected to be applicable, allowing
the energy dependence of the speed of light to be parameterized in a
model-independent way\cite{AMELINONATURE}. The photon speed $c'$ is written up
to second order in energy $E$ as:
\begin{equation}
c' = c ~ \left( 1 +  \xi \frac{E}{E_{\rm P}} + \zeta \frac{E^2}{
  E^2_{\rm P}}\right),
\label{eq:correction}
\end{equation}
where $\xi$ and $\zeta$ are free parameters. Even for the highest
photon energies currently measured the corrections are expected to be
very small. However, Amelino-Camelia et al. \cite{AMELINONATURE} suggested that these minuscule modifications
can add up to measurable time delays for photons from
cosmological sources. At a redshift $z$, simultaneously-emitted photons, with
energies $E_{1}$ and $E_{2}$, will
arrive at the observer with a time delay ${\rm\Delta} t = t_1 - t_2$  per energy
difference ${\rm\Delta E} = E_1 - E_2$ of \cite{PIRAN}:
\begin{equation}
\label{eq:linear}
 \frac{{\rm\Delta} t}{{\rm\Delta} E} \approx \frac{\xi}{ E_{\rm P} H_0} \int_0^z dz'
   \frac{(1+z')}{\sqrt{\Omega_m(1+z')^3 + \Omega_{\Lambda}}},
\end{equation}
where $\Omega_m$ = 0.3, $\Omega_{\Lambda}$ = 0.7 and $H_0$ = 70~km
s$^{-1}$ Mpc$^{-1}$ are the cosmological parameters as currently measured.
In the case of a vanishing linear term, the mean time delay of the photons per 
squared energy difference ${\rm\Delta} E^2 = E_1^2 - E_2^2$ is:
\begin{equation}
\label{eq:quad}
\frac{{\rm\Delta} t}{{\rm\Delta} E^2} \approx \frac{ 3 \zeta}{2 E^2_{\rm P} H_0}\int_0^z dz'
   \frac{(1+z')^2}{\sqrt{\Omega_m(1+z')^3 + \Omega_{\Lambda}}}
\end{equation}

The absence of such an energy dispersion has been used to set bounds on the
parameters $\xi$ and $\zeta$. Gamma-ray bursts and very high energies
flares of active galaxies have
been the primary targets of these ``time-of-flight'' studies.  
For the linear dispersion term in
Eq. \ref{eq:correction}, these measurements reach limits of
$|\xi| <$ 70--150 \cite{GRB2,GRBNOZ,GRB1,BOLMONT,GRBINTEGRAL,PIRAN2} for
gamma ray bursts.
For active galaxies, dispersion measurements exist for only two sources:
Mkn 421 and Mkn 501. Both are located at a similar redshift of
$\sim$0.03.
For Mkn 421, a limit of $|\xi| <$ 200 was set by the Whipple
collaboration during a flare in 1996 \cite{WHIPPLE}. For Mkn 501,
an indication of higher energy photons lagging the lower energy
ones was reported during a flare in 2005 by the
MAGIC collaboration \cite{MAGIC1}. This dispersion was recently
quantified to $|\xi| \sim$ 30 \cite{MAGIC2}. Since the signal is
however also marginally consistent with zero dispersion, limits of  $|\xi| <$
60 and $|\zeta| < 2.2 \times 10^{17}$ were derived
\cite{MAGIC2}. While limits in $\xi$ from time-of-flight measurements
are approaching unity and probing Planck-scale energies, limits on $\zeta$ are generally
still far from this domain. 

Time-of-flight measurements provide the most direct
and model-independent test of the constancy of the speed of light with
energy. However, alternative methods set more stringent limits
relying on additional assumptions: Limits
of  $|\xi| < 10^{-7}$ are deduced if the speed of light modifications in
Eq. \ref{eq:correction} are helicity dependent \cite{BIRE2,BIRE4}, as
predicted by some of the models \cite{LOOP1,EFT}. 
Also, constraining limits of $|\xi| <10^{-14}$  and $\zeta > -10^{-6}$ were
recently reported in \cite{SIGL} under several
assumptions, for example the sign of the speed of light
modification is assumed to be negative or helicity dependent
and standard kinematics are required to be valid in a Lorentz-violating regime.

\begin{figure}[h]
  \begin{center}
    \includegraphics[width=8cm]{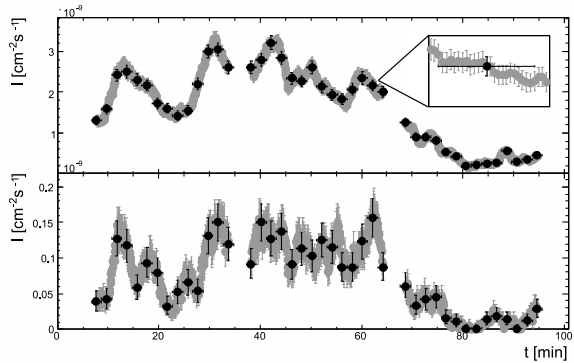}
  \end{center}
  \caption{Black points show the integral flux VHE light curves
  measured on July 28 from PKS~2155-304 by H.E.S.S. between 200-800 GeV (upper panel) and $>$800 GeV
  (lower panel), binned in two-minute time intervals. The zero time point is set to
  MJD 53944.02. Gray points show the oversampled light curve,
  for which the two-minute bins are shifted in units of
  five seconds. The inlay in the upper
  panel illustrates this in a zoom, where the horizontal error
  bar shows the duration of the bin in the original
  light curve.}
  \label{fig:lcs}
\end{figure}

A caveat of time-of-flight measurements
is that dispersion might be introduced by intrinsic source
effects, which could cancel out dispersion due to modifications of the speed of light.
In the case of a non-detection of dispersion this scenario is unlikely, since it requires
both effects to have the same time scale and opposite sign. However, this
``\textit{conspiracy of nature}''\cite{WHIPPLE}
can only be ruled out with certainty by observations of sources
at multiple distances, as -- in contrast to dispersion from speed of
light modifications -- source intrinsic dispersion should not scale
with distance.  Population studies of this kind have been performed
for gamma-ray bursts, resulting in limits of $|\xi| <$ 1300 \cite{FARAKOS,GRB1,BOLMONT,GRBINTEGRAL}.
For active galaxies the data-set is currently too sparse to perform
these studies.

In the present study, photon time delays were searched for during the
VHE flare of the active galaxy PKS 2155-304 observed by the High Energy Stereoscopic System (H.E.S.S.) on
July 28 in 2006. PKS~2155$-$304 is located at a redshift of z = 0.116 \cite{DIST}, almost four times more distant than Mkn 501 and
Mkn 421.
The light curve shows fast variability ($\sim$ 200~s)  and covers an energy range
of a few TeV with no significant spectral variability \cite{BIGFLARE}.
Considering the unprecedented photon
statistics ($\sim 10000$ photons) at these energies, this flare provides a
perfect testbed. The data presented here were analyzed using the
standard H.E.S.S. analysis, described in detail in \cite{CRAB}. Time
delays
between light curves of different energies were sought in order to
quantify a possible energy dispersion. For this, two different methods
were applied, which are described in the following.

\begin{figure}
  \begin{center}
    \includegraphics[width=8.5cm]{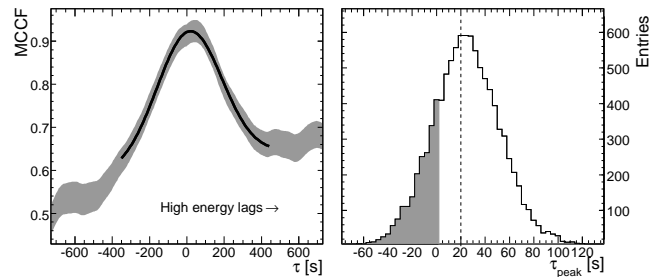}
  \end{center}
  \caption{Left: MCCF of the light curves in Fig. \ref{fig:lcs}. The black
  line shows the best fit of a Gaussian plus first degree
  polynomial. The peak of the fitted function is located at $\tau_{\rm
    peak} = 20$~s.
  Right: Cross Correlation Peak Distribution (CCPD)
  obtained from 10000 simulated light curves. The shaded area shows the range of the CCPD
  for $\tau_{\rm peak} \leq 0$, corresponding to 21\% of the total
  area.  The dotted line shows the position of $\tau_{\rm peak}$ from
  the left panel. The CCPD is slightly asymmetric, with a mean of 25~s and an RMS of 28~s}
  \label{fig:mdcf}
\end{figure}

The first method  determines the time lag between two light curves with the
 Modified Cross Correlation Function (MCCF) \cite{MCCF}. 
The MCCF is a standard cross correlation function \cite{DCF}, applied to
oversampled light curves. This allows time delays below the duration
 of the flux bins to be resolved \cite{MCCF}.
To optimize the energy gap between two energy bands,
while keeping good event statistics in both, the correlation analysis was performed on the light curves between
200 and 800 GeV and above 800 GeV (see Fig. \ref{fig:lcs}). The mean difference of the
photon energies between the two bands is 1.0 TeV and the mean quadratic difference
is 2.0 TeV$^2$. The MCCF of these light curves is shown in Fig. \ref{fig:mdcf}. In
order to measure the time delay, the  central peak of this distribution
 was fitted by a Gaussian function plus a first-degree polynomial,
resulting in a maximum at $\tau_{\rm peak}$ = 20~s.

The error on the measured time delay is determined by propagating the
flux errors via simulations.
Ten thousand simulated light curves were generated for each energy
band, by varying the flux points of the original oversampled light
curve within its measurement errors, taking into account bin
correlations. For each pair of light curves, the peak of the the MCCF was determined,
resulting in a Cross Correlation Peak Distribution (CCPD) shown in the
right panel of
Fig.  \ref{fig:mdcf}. The CCPD has an RMS of 28~s and yields the probability density
of the error of $\tau_{\rm peak}$ \cite{MAOZ,PETERSON}. For 21\% of the simulations
the time delay is negative, therefore the measured time delay of 20~s is not significantly
different from zero.

\begin{figure}
  \begin{center}
    \includegraphics[width=6.5cm]{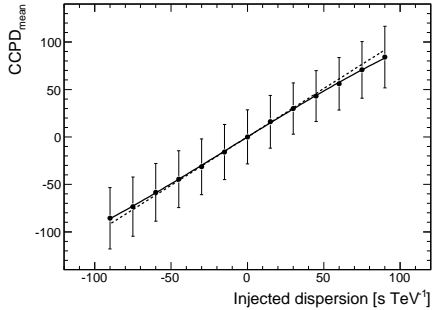}
  \end{center}
  \caption{Mean of the CCPD as a function of the
  dispersion injected in the H.E.S.S. data. The points have been
  shifted by the mean value of the CCPD of the original data
  shown in Fig.~\ref{fig:mdcf} to emphasize the relative time
  shifts. Each CCPD is derived from ten thousand simulated light curves.
  The error bars show the RMS of distributions. The solid line
  shows the calibration curve used to transform time shifts into
  dispersion. For comparison, the dotted line shows the linear response function expected
  from the mean energy of the correlated light curves (see text). }
  \label{fig:shifts}
\end{figure}

The response of the MCCF to energy dispersion is complex. Primarily,
dispersion is expected to shift light curves in time according to their mean
energy. However, dispersion also
broadens their structures and photons might even get
shifted out of a burst, decreasing the overall correlation.
These ``second order'' effects become increasingly
important once the time shifts approach the time
scale of the observed structures in the light curve.
The response of the MCCF to dispersion was
therefore determined by injecting artificial dispersion into the
H.E.S.S. data and measuring its effect on the CCPD.
As shown in Fig. \ref{fig:shifts}, the CCPD follows the injected time shift per energy
linearly in the range of interest here, confirming the expected behaviour. The second order effects mentioned only introduce small
deviations, visible at higher dispersion values.  Nevertheless, the measured time delays are transformed
to dispersion-per-energy with the calibration curve shown in
Fig. \ref{fig:shifts}, in order to take these effects into account. Since the measured $\tau_{\rm peak}$ was compatible
with zero, a 95\% confidence upper limit on a linear
dispersion of 73~s TeV$^{-1}$ is given. Applying the analogous procedure to a quadratic dispersion in
energy yields a 95\% confidence limit of 41~s~TeV$^{-2}$.

The accuracy of the MCCF method was verified with an
independent set of simulations. Eleven thousand new photon lists were
generated from the real data using a parametric bootstrap method. The
parametric  model was obtained from a polynomial spline
fit to the light curves in time bins of one minute and a fit of the
energy distribution of the events in the real data. The CCPD of these
new simulations confirmed the previously measured error on the time
delay. Artificially introduced dispersion was always recovered
within the expected accuracy. It should be noted that the dispersion limit 
does not depend strongly on the choice of preset parameters,
such as the energy ranges and time binning of the light curves and the fit
range of the MCCF peak. Varying these parameters within a
reasonable range has only a small effect ($\lesssim$5~s) on the final result. 

To confirm the result obtained with the MCCF analysis, the dispersion
measurement was repeated with an independent method, which is widely
applied in time lag studies of GRB light curves
\cite{GRBEXPLAIN,GRB1,BOLMONT}. Light curves were
constructed in two energy bands, and a search for extrema was done using a
Continuous Wavelet Transform (CWT) \cite{MALLAT}. For this the LastWave package
\cite{LASTWAVE} was employed, which provides a list of extrema candidates with their
positions. The extrema were associated in pairs between light curves
and their relative time delay
was measured. The association was performed with an algorithm based on the Lipschitz coefficient as in
\cite{GRB1,MALLAT}.

The two energy bands were chosen to be 210 to 250 GeV and above 600 GeV,
with a mean energy difference of 0.92 TeV. Since tiny
dispersions are to be probed, a time bin-width of 60 seconds was found to be optimal for
this study.  The CWT method identified two pairs of extrema with a mean time delay
of 27 seconds. In order to assess the error of this value,
samples composed of hundreds of Monte Carlo experiments were
analyzed for three linear dispersion values: 0 and $\pm$45 ~s
TeV$^{-1}$, in analogy to the MCCF calibration. The values of the
error on the measured time lag were found to range between 30 and 36
seconds. The relation between injected dispersion and measured time 
shift between light curves is again used to derive a limit on the dispersion, resulting in a
95$\%$ confidence limit of 100 s TeV$^{-1}$.  The impact of systematic
effects have also been investigated:
selection of gamma-like events and the choice of the energy domain or time binning of
the light curves change the results by 0.5$\sigma$ at most. Various
cuts on the CWT parameters have been applied and lead to negligible
changes in the extrema identification.

The measured limits on the energy dispersion translate into limits on the energy scale
of speed of light modifications. For a linear  dispersion in energy, Eq. \ref{eq:linear}
yields $|\xi| <$~17 (or $|\xi|^{-1}$~E$_p >$~7.2~$\times 10^{17}$ GeV) for the limit obtained with the MCCF
method, at 95\% confidence. The linear dispersion limits obtained from the Wavelet analysis
yields  a limit  of $|\xi| <$~23 (or $|\xi|^{-1}$~E$_p >$~5.2~$\times 10^{17}$ GeV),
confirming this result. These limits are the most constraining limits from time-of-flight
measurements to date. For a quadratic dispersion in energy, 
the MCCF method yields $|\zeta| <$~7.3~$\times 10^{19}$ (or
$|\zeta|^{-1/2}$~E$_p >$~1.4~$\times 10^{9}$ GeV) with
Eq. \ref{eq:quad}.

This measurement opens a new redshift range for population studies of
time delays from active galaxies, which are needed to rule out the
possibility of time delay cancellation. For a final verdict
on this question further VHE observations of active galaxies are
needed. However, the result already shows that the time delay reported for Mkn 501
in \cite{MAGIC2}, if considered significant, cannot be attributed to
speed of light modifications. Current and future 
instruments such as Fermi for gamma ray bursts, or the proposed Cherenkov
Telescope Array for
active galaxies, will further improve the sensitivity of
time-of-flight measurements, perhaps one day revealing deviations from
Einstein's postulate.
\smallskip
\begin{acknowledgments}
The support of the Namibian authorities and of the University of Namibia
in facilitating the construction and operation of H.E.S.S. is gratefully
acknowledged, as is the support by the German Ministry for Education and
Research (BMBF), the Max Planck Society, the French Ministry for Research,
the CNRS-IN2P3 and the Astroparticle Interdisciplinary Programme of the
CNRS, the U.K. Science and Technology Facilities Council (STFC),
the IPNP of the Charles University, the Polish Ministry of Science and
Higher Education, the South African Department of
Science and Technology and National Research Foundation, and by the
University of Namibia. We appreciate the excellent work of the technical
support staff in Berlin, Durham, Hamburg, Heidelberg, Palaiseau, Paris,
Saclay, and in Namibia in the construction and operation of the
equipment.
\end{acknowledgments}

\end{document}